\documentclass[aps,prb,amsmath,twocolumn,superscriptaddress,floatfix,footinbib,longbibliography]{revtex4}

\usepackage{graphicx}
\usepackage{epstopdf}

\usepackage[T1]{fontenc}
\usepackage[applemac]{inputenc}
\usepackage{lmodern}
\usepackage[english]{babel}
\usepackage{ae}
\usepackage{units}

\usepackage{amsmath,amssymb,natbib,bm}
\usepackage{psfrag}
\usepackage{subfigure}
\usepackage{amsthm}

\usepackage{fixmath}
\usepackage{booktabs}

\usepackage{slashed}

\usepackage[americaninductors]{circuitikz}
\usepackage{tikz}
\usetikzlibrary{arrows}

\usepackage{url}

\usepackage[colorlinks]{hyperref}

\usepackage{tabularx}

\hypersetup{%
        plainpages=true,
        breaklinks=true,
        hypertexnames=false,
        pageanchor=true,
        colorlinks=true,
        linkcolor={blue},
        citecolor={magenta},
        urlcolor={blue},
        anchorcolor={black}
      }

\renewcommand{\eqref}[1]{\mbox{Eq.~(\ref{#1})}}

\newcommand{\abs}[1]{\left|#1\right|}

\newcommand{\be}{\begin{equation}}
\newcommand{\ee}{\end{equation}}
\newcommand{\bea}{\begin{eqnarray}}
\newcommand{\eea}{\end{eqnarray}}

\begin{document}

\title{Coherent dynamics of a photon-dressed qubit}

\author{M.~P.~Liul}
\thanks{These authors contributed equally to this work}
\affiliation{B.~Verkin Institute for Low Temperature Physics and Engineering, Kharkov 61103, Ukraine}

\author{ C.-H.~Chien}
\thanks{These authors contributed equally to this work}
\affiliation{Department of Physics, National Tsing Hua University, Hsinchu 30013, Taiwan}

\author{C.-Y.~Chen}
\affiliation{Department of Physics, National Tsing Hua University, Hsinchu 30013, Taiwan}

\author{P.~Y.~Wen}
\affiliation{Department of Physics, National Chung Cheng University, Chiayi 621301, Taiwan}

\author{J.~C.~Chen}
\affiliation{Department of Physics, National Tsing Hua University, Hsinchu 30013, Taiwan}
\affiliation{Center for Quantum Technology, National Tsing Hua University, Hsinchu 30013, Taiwan}

\author{Y.-H.~Lin}
\affiliation{Department of Physics, National Tsing Hua University, Hsinchu 30013, Taiwan}
\affiliation{Center for Quantum Technology, National Tsing Hua University, Hsinchu 30013, Taiwan}

\author{S.~N.~Shevchenko}
\email[e-mail: ]{sshevchenko@ilt.kharkov.ua}
\affiliation{B.~Verkin Institute for Low Temperature Physics and Engineering, Kharkov 61103, Ukraine}

\author{Franco~Nori}
\affiliation{Theoretical Quantum Physics Laboratory, Cluster for Pioneering Research, RIKEN, Wakoshi, Saitama 351-0198}
\affiliation{Quantum Computing Center, RIKEN, Wakoshi, Saitama 351-0198, Japan}
\affiliation{Department of Physics, The University of Michigan, Ann Arbor, MI 48109-1040, USA}

\author{I.-C.~Hoi}
\email[e-mail: ]{iochoi@cityu.edu.hk}
\affiliation{Department of Physics, City University of Hong Kong,Tat Chee Avenue, Kowloon, Hong Kong SAR, China}
\affiliation{Department of Physics, National Tsing Hua University, Hsinchu 30013, Taiwan}

\date{\today}
\begin{abstract}
We consider the dynamics and stationary regime of a capacitively shunted transmon-type qubit in front of a mirror. The qubit is affected by probe and dressing signals. By varying the parameters of these signals and then analyzing the probe signal (reflected by the \textquotedblleft atom plus mirror\textquotedblright \ system), it is possible to explore the system dynamics, which can be described by the Bloch equation. The obtained time-dependent occupation probabilities are related to the experimentally measured reflection coefficient. The study of this type of dynamics opens up new horizons for better understanding the \textquotedblleft qubit plus mirror\textquotedblright \ circuit properties and the underlying physical processes, such as Landau-Zener-St\"{u}ckelberg-Majorana transitions.

\end{abstract}

\maketitle

\section{Introduction}
Topics related to quantum computing are attracting considerable attention [\onlinecite{Nielsen2010, Haffner2008, Arute2019, Buluta2011, Kjaergaard2020}]. One of the most promising building blocks of such devices are superconducting qubits (see, e.g., [\onlinecite{Kockum2019, Gu2017, Gambetta2017}]). These can be operated at nanosecond scales with millisecond coherence times [\onlinecite{Wang2022}], are controlled by microwaves and have lithographic scalability [\onlinecite{Oliver2013}]. Therefore, investigations of superconducting qubits could help in the development of quantum computers.

A superconducting qubit in a semi-infinite transmission line [\onlinecite{Hoi2015}] is important for quantum electrodynamics, especially waveguide quantum electrodynamics (WQED) [\onlinecite{Lalumiere13, Blais2020, Kannan2020}]. For example, in Ref.~[\onlinecite{wen18}] it was found that a transmon qubit embedded at the end of a transmission line can amplify a probe signal with an amplitude gain of up to 7\%;  while a single quantum dot [\onlinecite{Xu2007}] and natural atoms [\onlinecite{Wu1977}] show the signal amplifications at much lower levels: 0.005\% and 0.4\%, respectively. The investigation of our system can also address interesting physics issues in WQED, including: dynamics in atom-like mirrors [\onlinecite{Mirhosseini2019}], collective Lamb shift [\onlinecite{Wen2019}], generation of non-classical microwaves [\onlinecite{Hoi2012}], the dynamical Casimir effect [\onlinecite{Wilson2011}], cross-Kerr effect [\onlinecite{Hoi2013a}], photon routing [\onlinecite{Hoi2011}], probabilistic motional averaging [\onlinecite{Karpov2019}], etc.

Driven quantum systems can be described in terms of Landau-Zener-Stuckelberg-Majorana (LZSM) transitions [\onlinecite{Oliver2005, Sillanpaa2006, Shevchenko2010, Ivakhnenko2022}]. If driven periodically, these experience interference. The corresponding LZSM interferometry is important both for studying fundamental quantum phenomena and as a convenient tool for characterizing quantum systems. The use of LZSM interferometry for controlling quantum dynamics was studied in Refs.~[\onlinecite{Ivakhnenko2022, Gorelik1998, Wu19, Shevchenko, Kofman2022, Liul2022, Satanin2012}]. Quantum logic gates can also be implemented using LZSM dynamics [\onlinecite{Campbell2020}].    

In a preceding work, Ref.~[\onlinecite{Wen2020}], we explored LZSM interferometry spectroscopically, i.e. in the frequency domain. Taking advantages of the strong coupling between propagating fields and qubits, as well as the ease of fabrication, circuits with superconducting qubits in front of a mirror provide a versatile platform to study the dynamics of LZSM interference, as compared with other quantum two-level systems.

The rest of the paper is organized as follows. Section~II is devoted to the description of the experiment. In Sec.~III the theoretical aspects of the problem are described; we introduce the Hamiltonian of the system and the equation of motion which was solved to obtain the quantities shown. Section~IV presents our results: a comparison of the theory and the experiment is given; the general patterns of the system are described and explained; additional results are given in Appendix A. In Sec.~V we present our conclusions.     
\begin{figure*}
	\includegraphics[width=0.95 \linewidth]{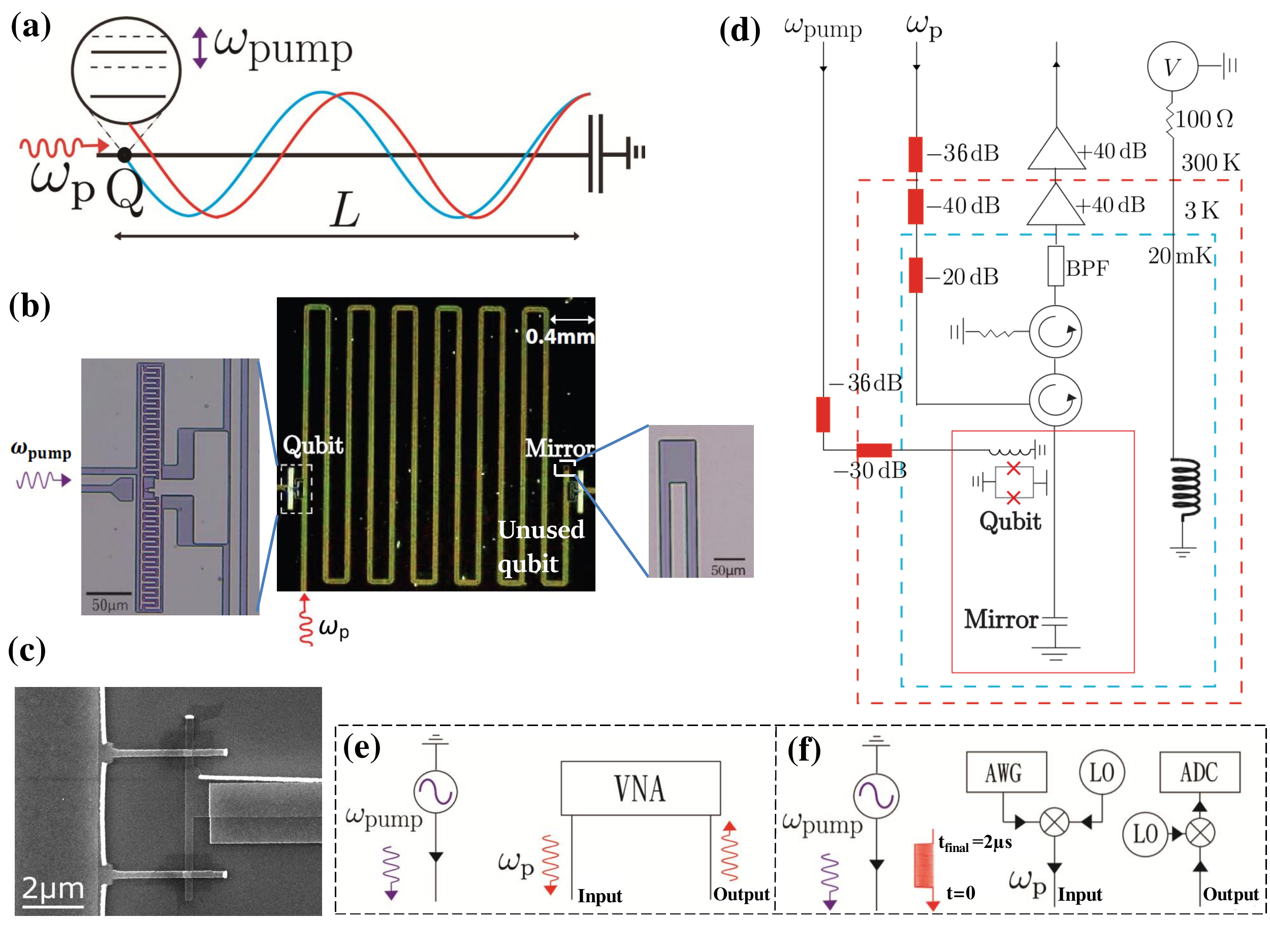}
	\caption{Device and its characterization. (a)~Conceptual sketch of the device: a two-level atom, point-like object (denoted by Q), is coupled to a semi-infinite transmission line waveguide. The atom is located at a distance $L\simeq 33$ mm (1.84$\lambda$, indicated by the red curve, 1.75$\lambda_{\rm node}$, indicated by the blue curve) away from the mirror (capacitance). A pump tone with frequency $\omega _{\mathrm{pump}}$ is applied to modulate the transition frequency of the two-level atom. A weak probe tone with frequency $\omega _{\mathrm{p}}$ is applied to the atom-mirror system to measure the reflection coefficient. (b)~Micrographs of the device. The magnification of the active qubit is shown to the left, where the superconducting qubit is intentionally designed to be weakly coupled to the transmission line. The weak coupling enables us to measure the temporal dynamics with a nano-second digitizer. This is the main difference between this work and a previous one [\onlinecite{Wen2020}] which focused on the stationary regime. A capacitor to ground, which creates an antinode voltage field at the end of the transmission line and acts as a mirror, is shown to the right. The transmission line ends in another qubit, which is designed for another experiment and is not participating in the experiment (because it is far detuned). (c)~Contains a scanning-electron-microscope picture of the qubit dc-SQUID which enables flux tunability. (d)~The experimental setup shows the probe tone and pump tone being applied to the atom-mirror system. Panels (e) Vector Network Analyzer [VNA], Signal Generator and (f) Arbitrary Waveform Generator [AWG], Local Oscillator[LO], Analog-to-Digital Converter [ADC] show the setup for Figs.~\ref{fig_calibration},~\ref{Fig_interferogram} and Figs.~\ref{Fig_time_f_pump},~\ref{Fig_time_line},~\ref{Fig_time_delta},~\ref{Fig_no_flux} respectively.  
		\label{fig_device}}
\end{figure*}
\section{Experiment}
Our device consists of a transmon qubit, embedded at a distance ($L\simeq 33$ mm, where the resonant frequency corresponds to 1.84$\lambda$) of a finite quasi-one-dimensional transmission line with characteristic impedance $Z_0 \simeq \unit[50]{\Omega}$, as shown in Fig.~\ref{fig_device}~(a,b). The transmission line allows the formation of standing EM waves along the transmission line, therefore, the voltage field strength experienced by the qubit can be controlled by the location of the qubit in the transmission line, as illustrated in Fig.~\ref{fig_device}(a). In principle, we could have used a short instead of an open end for the transmission line. However, this changes the boundary conditions. In particular, the phase of the incident wave acquires a $\pi$ phase shift, instead of a zero phase shift. And the voltage at the end of transmission line would be at the node instead of at the antinode. Figure~\ref{fig_device}~(c) contains a scanning-electron-microscope picture of the qubit dc-SQUID which enables flux tunability. We only consider the lowest two energy levels of the transmon, and neglect all the higher levels. The energy splitting of the two-level atom is $\hbar \omega_{10}(\rm \Phi) \approx \sqrt{8\textit{E}_{\rm J}(\rm \Phi) \textit{E}_{\rm C}} - \textit{E}_{\rm C}$, with charging energy $E_{\rm C}$ (which is approximately equal to the anharmonicity), $E_{\rm C} = e^2 / 2C_\Sigma$, where $e$ is the elementary charge and $C_\Sigma$ is the total capacitance of the transmon, and the Josephson energy $E_{\rm J}$, which can be tuned by the external magnetic flux $\rm \Phi$ of a magnetic coil. The detailed measurement setup is presented in Fig.~\ref{fig_device}(d). Here $\omega_{\rm p}$ is the probing frequency (indicated in red color in Fig.~\ref{fig_device}(b,e,f)), a continuous wave created by the vector network analyzer or the microwave pulse from the arbitrary wave generator; we input the continuous wave with pump frequency $\omega_{\rm pump}$ by the RF source (indicated by purple color in Fig.~\ref{fig_device}(b,e,f)). 

\begin{table*}
\begin{tabular}{ |p{4cm}|p{4cm}|p{4cm}|p{4cm}|}
\hline
Parameter& Description &Value in Ref.~[\onlinecite{Wen2020}]&Current value\\
\hline
$\omega_{\rm node}$   & Node frequency    &$2\pi\times4.75~\mathrm{GHz}$&$2\pi\times4.38~\mathrm{GHz}$\\
$\omega_{10}$   & Qubit frequency    &$\simeq  \omega_{\rm node}$&$2\pi\times5~\mathrm{GHz}$\\
$\delta$&Pump amplitude &$\sim 2\pi\times0.1~\mathrm{GHz}$&$<2\pi\times 40~\mathrm{MHz}$\\
$\omega_{\rm pump}$ &Pump frequency&$< 2\pi\times0.1~\mathrm{GHz}$&$2\pi\times 1 \sim 15~\mathrm{MHz}$\\
$\omega_{\rm p}$&Probe frequency&$\simeq \omega_{\rm node}$&$2\pi\times5~\mathrm{GHz}$\\
$\mathrm{\Gamma}_{1}$&Relaxation rate&$<2\pi\times 5~\mathrm{MHz}$&$2\pi\times0.28~\mathrm{MHz}$\\
$\mathrm{\Gamma}_{\phi}$&Pure dephasing rate&$\sim 2\pi\times 3~\mathrm{MHz}$&$2\pi\times0.61~\mathrm{MHz}$\\
$\mathrm{\Gamma}_{2}=\mathrm{\Gamma}_{1}/2 + \mathrm{\Gamma}_{\phi}$&Decoherence rate&$\sim 2\pi\times 5.5~\mathrm{MHz}$&$2\pi\times0.75~\mathrm{MHz}$\\
\hline
\end{tabular}
\caption{\label{table} Comparison of parameters between this work and a previous one, Ref.~[\onlinecite{Wen2020}].}
\end{table*}

\subsection{Characterizing the qubit by single-tone scattering with a weak probe}
\begin{figure}
	\includegraphics[width=1 \linewidth]{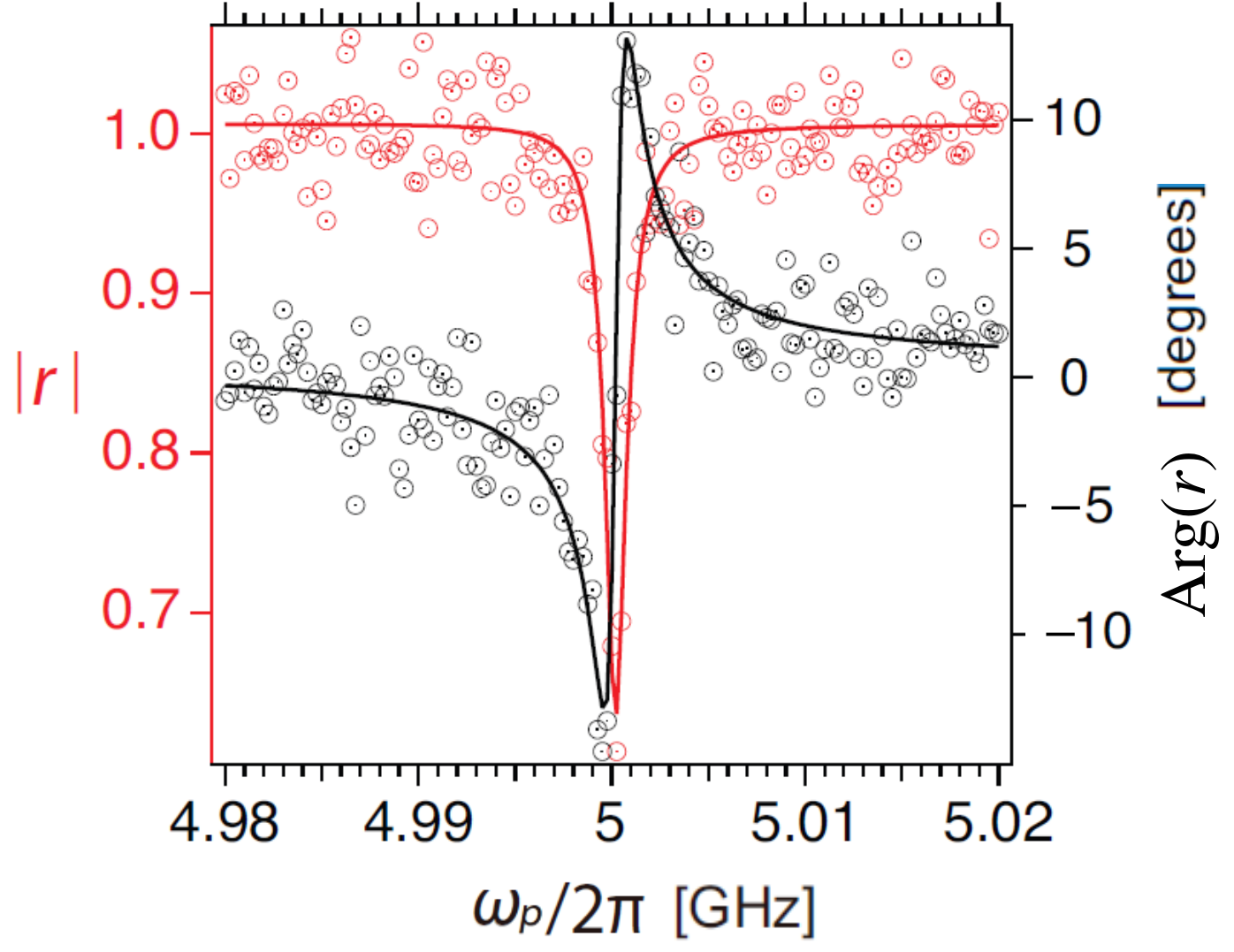}
	\caption{Reflection coefficient (magnitude response in red and phase response in black) as a function of probe frequency $\omega_{\mathrm{p}}$ for a weak probe ($-$166\,dBm), see Fig.~\ref{fig_device}(e) with pump off. Data are shown by circles and the solid curves are the fit, using a similar fitting method as in Ref.~[\onlinecite{Lin2020}].
		\label{fig_calibration}}
\end{figure}
We characterize our qubit by a single-tone scattering with a weak probe, see Fig.~\ref{fig_calibration}. Fitting the magnitude and phase response of the reflection coefficient $r$ by using the circle fit equation Eq.~(\ref{fitting_eq}) [\onlinecite{Probst2015}] one could extract the resonant frequency $\omega_{10}$, the relaxation rate $\rm \Gamma_1$ and decoherence rate $\rm \Gamma_2$:  
\begin{eqnarray}
r = 1 - \frac{\mathrm{\Gamma_{1}}}{\mathrm{\Gamma_{2}} + i(\omega_{\mathrm{p}} - \omega_{10})},
\label{fitting_eq}
\end{eqnarray}
here $r$ is defined as the reflected field amplitude divided by the incoming field amplitude. The extracted values are the following: $\omega_{10}/2\pi~=~\unit[5]{GHz}$,  $\mathrm{\Gamma}_{1}/2\pi=\unit[0.28]{MHz}$, and $\mathrm{\Gamma}_{2}/2\pi=\unit[0.75]{MHz}$. These parameters will be used in the theory later. Note that we assume the incoming amplitude to be the same as the reflected amplitude when the qubit is detuned. By using two-tone spectroscopy, we know that $E_{\rm C}/h$~(anharmonicity)~=~$\unit[220]{MHz}$ (data not shown), which is much larger than any Rabi frequency in this work, and our two-level atom assumption is valid. From $\omega_{10}$ and $E_{\rm C}$, we know that $E_{\rm J}/h=\unit[15.5]{GHz}$. 
\begin{figure*}
	\includegraphics[width=1 \linewidth]{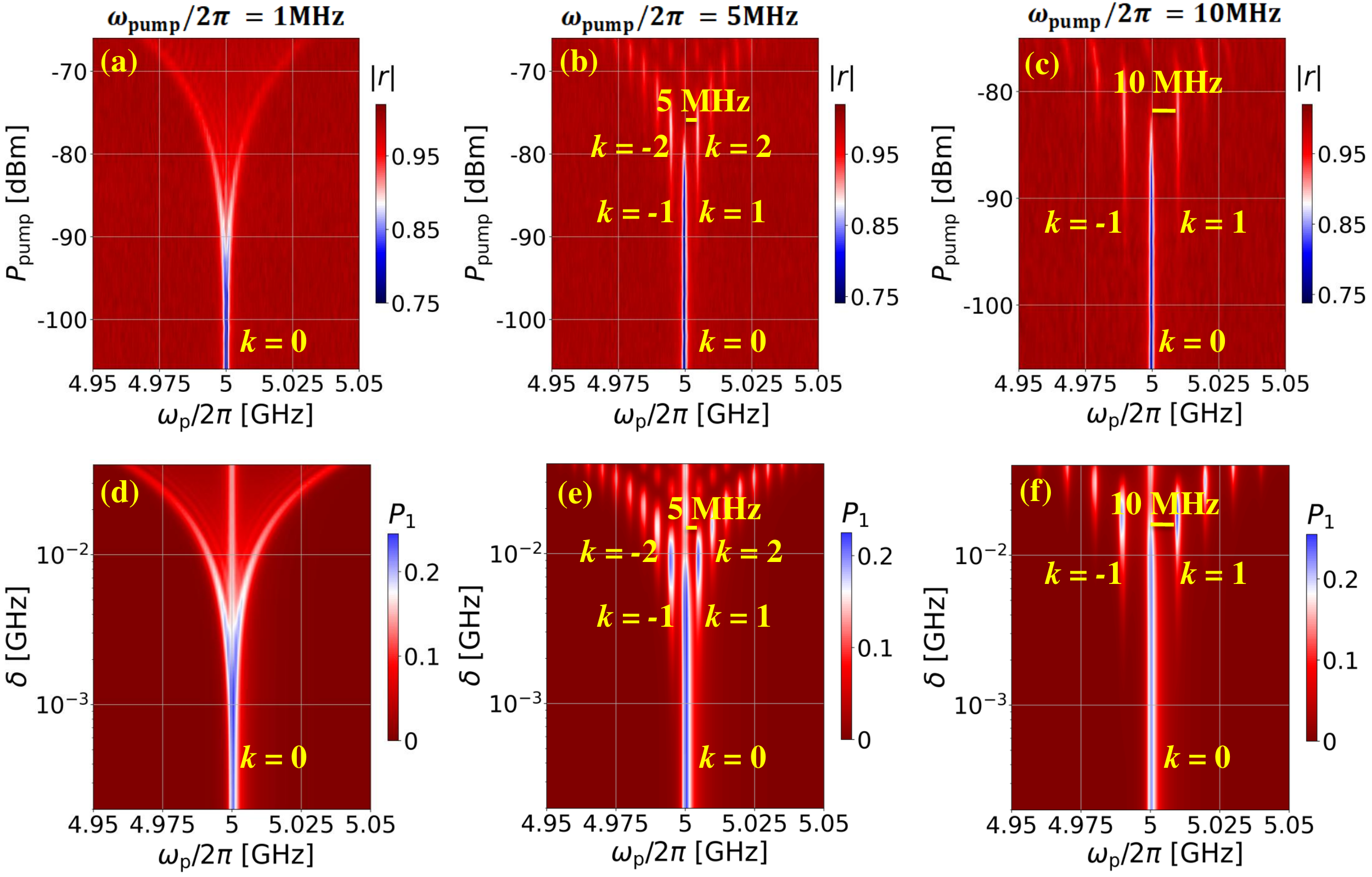}
	\caption{LZSM interferograms. These are shown via the dependence of the reflection coefficient $\abs{r}$ on the pump power $P_{\rm pump}$ and probe frequency $\omega_{\rm p}$ at fixed pump frequency $\omega_{\rm pump}$ for a weak probe $P_{\rm p}=\unit[-152]{dBm}$ in the top panels (a,b,c) for the experimental measurements. While the bottom panels (d,e,f) show the theoretically calculated upper-level occupation probability $P_{1}$ as a function of the probe frequency $\omega_{\rm p}$ and the pump amplitude $\delta$ for $G(\omega_{\rm p}/2\pi~=~5~\mathrm{GHz})~=~2\pi\times0.7~\mathrm{MHz}$. The qubit is irradiated by a pump with frequency (a)~$\omega_{\rm pump}/2\pi=\unit[1]{MHz}$, (b)~$\omega_{\rm pump}/2\pi=\unit[5]{MHz}$, (c)~$\omega_{\rm pump}/2\pi=\unit[10]{MHz}$. In (c), the drift on multi-photon resonance at high power is due to flux drift.
		\label{Fig_interferogram}}
\end{figure*}

\begin{figure*}
	\includegraphics[width=0.95 \linewidth]{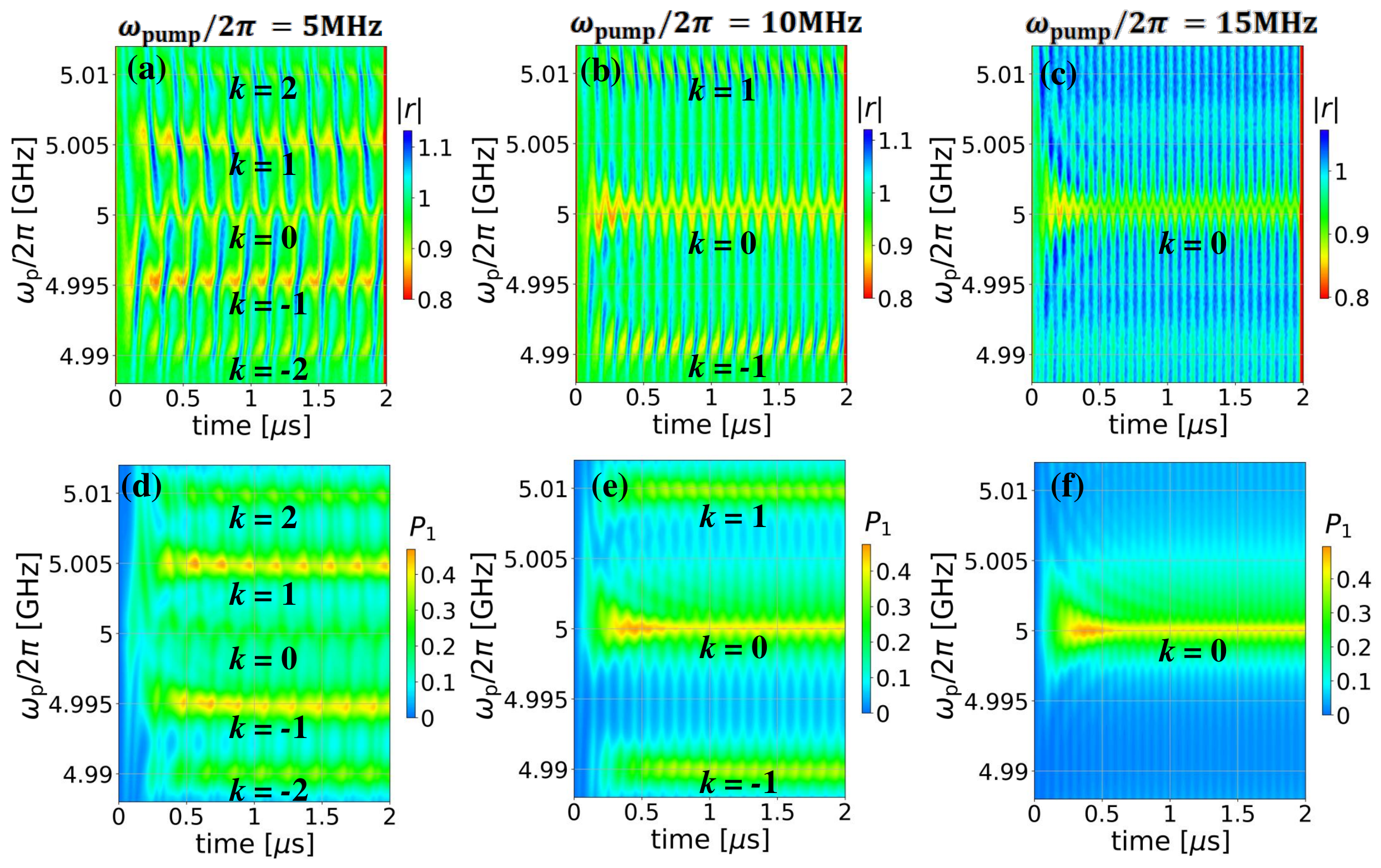}
	\caption{Coherent dynamics of the transmon qubit: dependence of the reflection coefficient $\abs{r}$ (the upper-level occupation probability $P_{1}$) using the probe power $P_{\rm p}=\unit[-146]{dBm}$ [$G(\omega_{\rm p}/2\pi~=~5~\mathrm{GHz}) = 2\pi\times1.4~\mathrm{MHz}$] and the pump power $P_{\rm pump}=\unit[-78.5]{dBm}$ ($\delta=\unit[10]{MHz}$) on the probe frequency $\omega_{\rm p}$ and time $t$. Plots (a,b,c) present experimental results, (d,e,f) show plots built by data computed theoretically. The qubit is irradiated by a pump with frequency (a) $\omega_{\rm pump}/2\pi=\unit[5]{MHz}$, (b) $\omega_{\rm pump}/2\pi=\unit[10]{MHz}$, (c)~$\omega_{\rm pump}/2\pi=\unit[15]{MHz}$. Panels (d-f)  show the corresponding data computed theoretically for the qubit upper-level occupation probabilities $P_{1}$.
		\label{Fig_time_f_pump}}
\end{figure*}
\subsection{Reflection coefficient versus pump power and probe frequency}
We apply both probe and dressing (pump) signals to the transmission line and the on-chip flux line (which modulates the transition frequency of the qubit), respectively. We then measure the reflection coefficient $r$ in the (steady state) frequency domain, see Fig.~\ref{fig_device}(e). Both frequency and power for the pump tone and probe tone are all tunable. Here, we scan the frequency of probe $\omega_{\rm p}$, frequency of pump $\omega_{\rm pump}$, power of pump $P_{\rm pump}$, and measure the reflection coefficient $\abs{r}$ with a fixed power of the weak probe tone. 

The results are shown in Fig.~\ref{Fig_interferogram}. In each subplot of Fig.~\ref{Fig_interferogram}, we fix the pump frequency, vary the pump power ($y$-axis) and probe frequency ($x$-axis). We see a LZSM interferometry pattern, where we clearly observe multi-photon resonances, which occur at $\omega_{\rm p}-\omega_{10}\equiv \mathrm{\Delta} \omega=k\omega_{\rm pump}$, where $k$ is an integer number. From the experimental and theory plots, one can calibrate the Rabi frequency and the pump power. In addition, from these plots it is possible to extract qubit parameters, such as the relaxation rate $\mathrm{\Gamma}_1$, pure dephasing rate $\mathrm{\Gamma}_{\phi}$, and decoherence rate $\mathrm{\Gamma}_2$. The parameters obtained are shown in Table~\ref{table}. Note that the $k=0$ transition is observed because we tune the qubit frequency away from the node frequency, similar to Fig.~5(c) in Ref.~[\onlinecite{Wen2020}]. If the qubit is at a node then a transition for $k=0$ is not observed (see Fig.~5(b) in Ref.~[\onlinecite{Wen2020}]).

As shown in Fig.~\ref{Fig_interferogram}(b), the higher pump powers allow to resolve more sidebands, visible there up to $k=\pm5$. The experimental LZSM interferometry pattern shown in Fig.~\ref{Fig_interferogram}(a,b,c) matches very well with the theory in Fig.~\ref{Fig_interferogram}(d,e,f). Also one can see the drift on the multi-photon resonance at high power in (c), which is due to flux drift (caused by environmental background flux), because the resonance frequency is controlled by the flux. Detailed calculations are shown in Sec.~III. Since we are using a weak probe, the upper-level occupation probability $P_{1}$ is low. 

\subsection{LZSM interferometry of the system in the stationary regime}
In this subsection we give a brief comparison with previous related research, in Ref.~[\onlinecite{Wen2020}]. The main difference between these two studies is that in the previous experiment the qubit was located in the node [blue curve in Fig.~\ref{fig_device}(a)], thus it was "hidden" or "decoupled" from the transmission line. In other words, the qubit was exposed to the electric field, but could not experience the electric field because the qubit was located at the node. By tuning the qubit to the node frequency, we could decouple the $k=0$ sideband and see the other sidebands. 

In this work, we tuned slightly away from the node frequency [red curve in Fig.~\ref{fig_device}(a)], and we see the time dynamics of the  $k= 0$ photon-dressed resonance. The advantage of shifting slightly away from the node frequency is in obtaining a long coherence time $T_{2}=1/\Gamma_{2} \sim 1/2\pi \times \unit[0.75]{MHz}\sim \unit[212]{ns}$ (see Table~\ref{table}), which is important to observe the time dynamics of the photon-dressed resonance, using a finite-time-resolution digitizer. 

The time domain measurements were not made in Ref.~[\onlinecite{Wen2020}], because the relaxation rate and pure dephasing rate were much higher than in the current work, where the qubit-transmission line coupling is intentionally designed to be weak. Regarding the relaxation time and pure dephasing time in Ref.~[\onlinecite{Wen2020}], we could not resolve these with a finite bandwidth digitizer ($\unit[5]{ns}$ resolution), because the dynamics was too fast.

Also, compared to the preceding work, the coupling capacitor to the transmission line is decreased because of the intended and desired weak coupling, where the relaxation rate is smaller and the coherence time is larger. This allows us to reveal the dynamics with a nano-second digitizer. In order to keep the same charging energy, which determines the anharmonicity, we have to increase the shunt capacitance to keep the same total capacitance.

\subsection{Temporal dynamics of the atom-mirror under both pump and probe signals}
We study the time dynamics of the atom-mirror system under both probe and dressing (pump) signals. In particular, we send a probe square pulse (Gaussian rise $\sim10$\,ns) to the transmission line and continuous sinusoidal wave pump to the on-chip flux line, see Fig.~\ref{fig_device}(f). We measure the reflection coefficient as a function of time and probe the frequency for a weak probe under the influence of a fixed pump power and fixed pump frequency. 
\begin{figure*}
	\includegraphics[width=0.95 \linewidth]{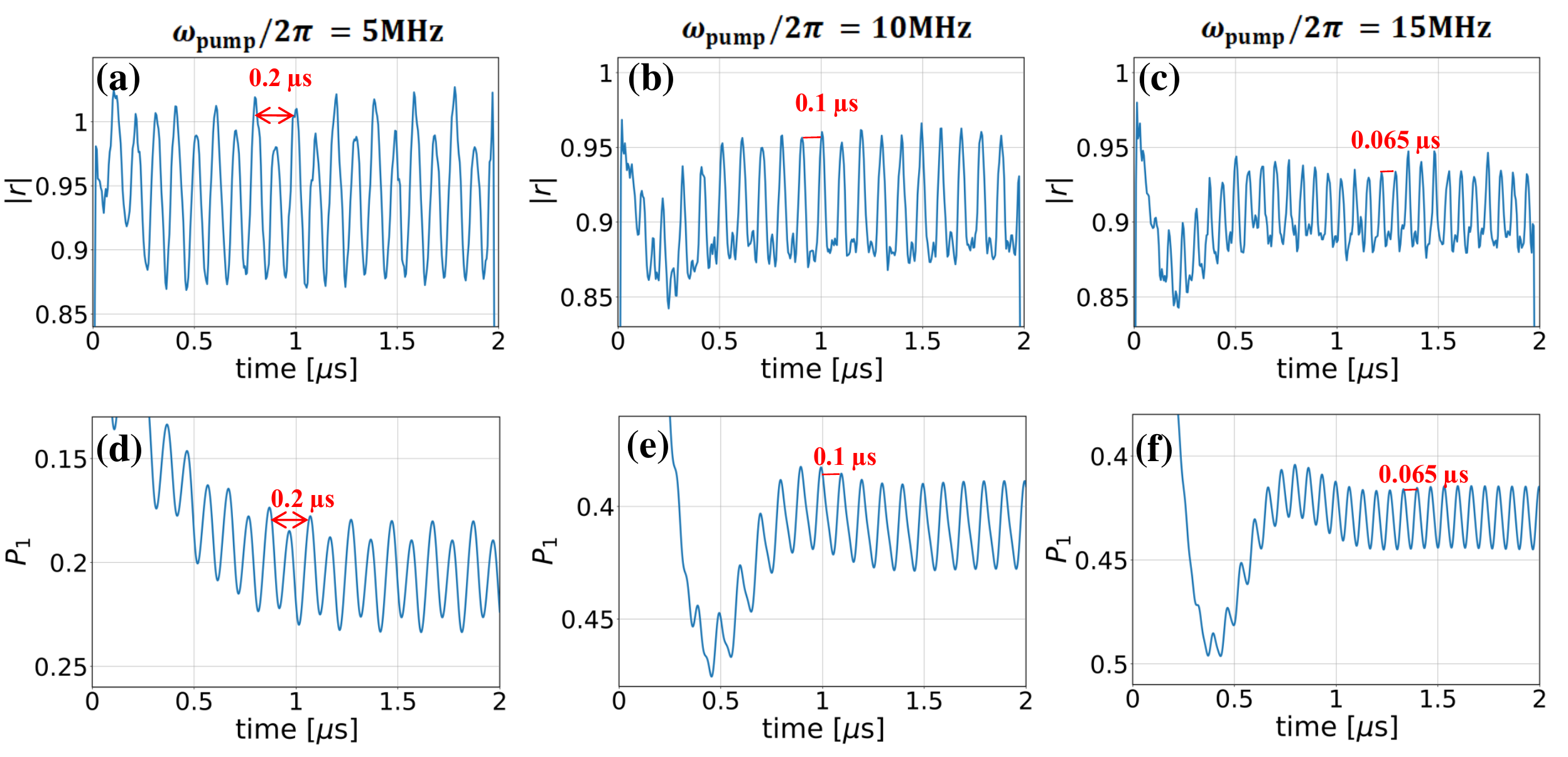}
	\caption{Line cut of Fig.~\ref{Fig_time_f_pump} along $\omega_{\rm p}/2\pi=\unit[5]{GHz}$. Coherent dynamics of the transmon qubit: dependence of the reflection coefficient $\abs{r}$ (the upper-level occupation probability $P_{1}$) on the time $t$ at fixed pump frequency $\omega_{\rm pump}$ with $\omega_{\rm p}/2\pi=\unit[5]{GHz}$, $P_{\rm p}=\unit[-146]{dBm}$ [$G(\omega_{\rm p}/2\pi~=~5~\mathrm{GHz}) = 2\pi\times1.4~\mathrm{MHz}$], $P_{\rm pump}=\unit[-78.5]{dBm}$ ($\delta=\unit[10]{MHz}$). Plots (a,b,c) present experimental results, while panels (d,e,f) show plots computed theoretically. The qubit is irradiated by a signal with frequency (a)~$\omega_{\rm pump}/2\pi=\unit[5]{MHz}$, (b)~$\omega_{\rm pump}/2\pi=\unit[10]{MHz}$, (c)~$\omega_{\rm pump}/2\pi=\unit[15]{MHz}$. Panels (d,e,f)  show the corresponding data computed theoretically for the qubit upper-level occupation probabilities $P_{1}$. For a clearer comparison between theory and experiments, the $y$-axis of the theoretical plots in (d,e,f) were cropped and inverted. 
		\label{Fig_time_line}}
\end{figure*}

In Fig.~\ref{Fig_time_f_pump}, each plot is taken at a fixed pump power $P_{\rm pump}=\unit[-78.5]{dBm}$ and pump frequency. Note that $\omega_{\rm pump}/2\pi=\unit[5]{MHz}$ for panels (a,d); $\omega_{\rm pump}/2\pi=\unit[10]{MHz}$ for panels (b, e); and $\omega_{\rm pump}/2\pi=\unit[15]{MHz}$ for panels (c, f). The probe pulse starts at the beginning of the plot at $t=0$. In Fig.~\ref{Fig_time_f_pump} (a,b,c) the reflection coefficient reveals a transient dynamics starting at $t=0$. This transient dynamics, affected by the initial conditions, ends up in a stationary solution, determined by the competition of driving and relaxation. In addition, we see the time dynamics of the multi-photon resonances, which occur at  $\mathrm{\Delta} \omega=k\omega_{\rm pump}$, labeled by $k=-2, -1, 0, +1, +2$. And the multi-photon resonances are slightly asymmetric around $k=0$. All of these features are consistent with the theory described in Sec.~III. 

In Fig.~\ref{Fig_time_line}, by taking line cuts of Fig.~\ref{Fig_time_f_pump}, we show detailed features of the transient dynamics at various fixed pump frequencies $\omega_{\rm pump}$. Moreover, in Fig.~\ref{Fig_time_delta}, we also fix the pump frequency to 10\,MHz and vary the pump power in (a,b,c). We see the multi-photon resonances, occurring at $k=-1,+1$, becoming weaker and weaker from (a) to (c), as the pump power decreases. 

In Fig.~\ref{Fig_time_line}, we can see the line cut along Fig.~\ref{Fig_time_f_pump} at $\omega_{\rm p}/2\pi=\unit[5]{GHz}$. For a clearer comparison between theory and experiment, the $y$-axis for the theoretical plots was cropped and inverted. We see the transient dynamics (oscillations with a frequency inversely proportional to the pump frequency $\omega_{\rm pump}$) around $T_2=1/\mathrm{\Gamma}_2\sim\unit[212]{ns}$ for $\omega_{\rm pump}/2\pi=\unit[15]{MHz}$, where $T_2>2\pi/\omega_{\rm pump}$ in Fig.~\ref{Fig_time_line}(c). When $T_2\sim2\pi/\omega_{\rm pump}$, the transient dynamics is not clear, as shown in Fig.~\ref{Fig_time_line}(a). In the steady state, the period of oscillations is the inverse of the pump frequency, as expected. The theory plots show the upper-level occupation probability $P_{1}$, where the transient dynamics is around $T_1=1/\mathrm{\Gamma}_1\sim\unit[568]{ns}$. Also, for a better understanding of the qubit dynamics formation driven by the flux pump, we show the case when pump is off. From this, we learn that the features at $k~=~\pm 1$ disappear. The corresponding plots are presented in Fig.~\ref{Fig_no_flux} in the Appendix. 

\section{Theoretical description}
\begin{figure*}
	\includegraphics[width=0.95 \linewidth]{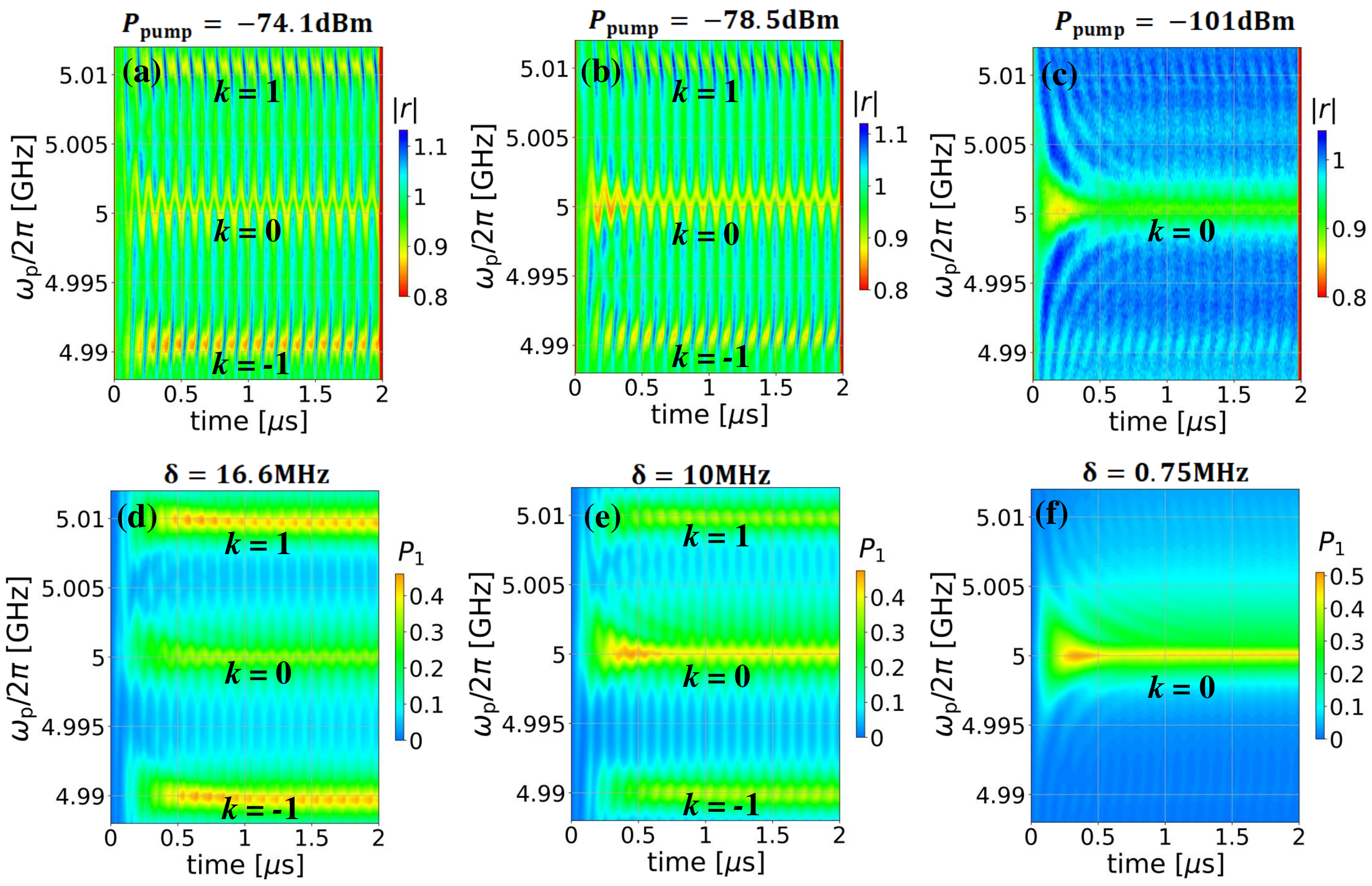}
	\caption{Coherent dynamics of the transmon qubit: dependence of the reflection coefficient $\abs{r}$ (the upper-level occupation probability $P_{1}$) versus time $t$, using the probe power $P_{\rm p}=\unit[-146]{dBm}$ [$G(\omega_{\rm p}/2\pi~=~5~\mathrm{GHz}) = 2\pi\times1.4~\mathrm{MHz}$ for the theory] on the probe frequency $\omega_{\rm p}/2\pi$. Plots (a,b,c) present experimental results; while (d,e,f) show plots built by data computed theoretically. The qubit is irradiated by a pump signal with frequency $\omega_{\rm pump}/2\pi=\unit[10]{MHz}$ and power (a)~$P_{\rm pump}=\unit[-74.1]{dBm}$, (b)~$P_{\rm pump}=\unit[-78.5]{dBm}$, (c)~$P_{\rm pump}=\unit[-101]{dBm}$. The theoretical results from the qubit upper-level occupation probability $P_{1}$ are shown for $\delta=\unit[16.6]{MHz}$ in (d), for $\delta=\unit[10]{MHz}$ in (e), and for $\delta=\unit[0.75]{MHz}$ in (f).
		\label{Fig_time_delta}}
\end{figure*}

In Ref.~[\onlinecite{Wen2020}], the experimentally measured reflection coefficient $\abs{r}$ is associated with the theoretically calculated probability of an upper level occupation $P_{1}$ (increasing $P_{1}$ corresponds to decreasing $\abs{r}$). The computations were done in the diabatic (charge) basis. Here we use the same correspondence between theory and experiment and make our calculations in the diabatic basis. The system can be described by the Hamiltonian:%
\begin{equation}
H=- \frac{B_{z}}{2}\sigma _{z}-\frac{B_{x}}{2}\sigma _{x},
\label{Hamiltonian}
\end{equation}%
where the diagonal part corresponds to the energy-level modulation 
\begin{equation}
B_{z}/\hbar =\omega _{10}+\delta \sin \omega _{\mathrm{pump}}t,  \label{Bz}
\end{equation}%
the off-diagonal part characterizes the coupling to the probe signal%
\begin{equation}
B_{x}/\hbar =G\sin \omega _{\mathrm{p}}t.  \label{Bx}
\end{equation}%
To remove the fast driving from the Hamiltonian, Ref.~[\onlinecite{Wen2020}] considered the unitary transformation $U=\exp \left( -i\omega _{\mathrm{p}}\sigma _{z}t/2\right)$
and the rotating-wave approximation [\onlinecite{Silveri15}, \onlinecite{Ono2020}] to obtain the new Hamiltonian
\begin{equation}
H_{1}=-\frac{\hbar \widetilde{\Delta \omega }}{2}\sigma _{z}+\frac{\hbar G%
}{2}\sigma _{x},  
\label{H_in_RWA}
\end{equation}%
where%
\begin{eqnarray}
\widetilde{\Delta \omega } &=&\Delta \omega +f(t), \\
\Delta \omega  &=&\omega _{\mathrm{p}}-\omega _{\mathrm{10}}, \\
f(t) &=&\delta \sin \omega _{\mathrm{pump}}t.
\end{eqnarray}%
Here $\delta$ is the amplitude of the energy-level modulation, $G$ characterizes the coupling to the probe
signal (Rabi frequency of the probe signal). According to Ref.~[\onlinecite{Wen2020}]
\begin{eqnarray}
G=\frac{\omega_{\mathrm{p}} - \omega_{\mathrm{node}}}{\omega_{\mathrm{node}}}G_{0},
\label{G_expression}
\end{eqnarray}
where $\omega_{\mathrm{node}}$ describes the qubit position in a semi-infinite transmission line [corresponding to the blue curve in Fig.~\ref{fig_device}(a)], and $G_{0}$ is proportional to the probe signal amplitude. Such a dependence causes the asymmetry about the line $\omega_{\rm p}/2\pi = \unit[5]{GHz}$ in Figs.~\ref{Fig_interferogram},~\ref{Fig_time_f_pump},~\ref{Fig_time_delta}. For the current experiment $\omega_{\mathrm{node}}/2\pi=\unit[4.38]{GHz}$. For any multi-photon resonance close to the node frequency, the linewidth will be narrower. The closer it is, the narrower it is. Therefore, it gives the asymmetry about the qubit resonance at $\omega_{\rm p}/2\pi = \unit[5]{GHz}$. Moreover, one can see that if $\omega_{\mathrm{p}} = \omega_{\mathrm{node}}$ the qubit is \textquotedblleft hidden\textquotedblright\ or
\textquotedblleft decoupled\textquotedblright\ from the transmission line, with $G=0$.

In order to describe the qubit dynamics, we use the Lindblad equation, which in the diabatic basis with the Hamiltonian~(\ref{H_in_RWA}) has the form:%
\begin{eqnarray}
\frac{d\rho}{d t} = -\frac{i}{\hbar}\left [ \widehat{H}_{1}, \rho \right ] + \sum_{\alpha}\breve{L}_{\alpha}\left [ \rho \right ],
\label{Bloch_eq}
\end{eqnarray}
where $\rho = \begin{pmatrix} \rho_{00} & \rho_{01} \\ \rho^{*}_{01} & 1 - \rho_{00} \end{pmatrix}$ is the density matrix, such that $P_{1} = 1 - \rho_{00}$. The Lindblad superoperator $\breve{L}_{\alpha}$ characterizes the system relaxation caused by interactions with the environment,
\begin{eqnarray}
\breve{L}_{\alpha}\left [ \rho \right ] = L_{\alpha}\rho L_{\alpha}^{+} - \frac{1}{2}\left\{L_{\alpha}^{+}L_{\alpha}, \rho \right\},
\label{Lindblad_superoperator}
\end{eqnarray}
where $\left\{a,b \right\} = ab + ba$ is the anticommutator. For a qubit there are two possible relaxation channels: energy relaxation (described by $L_{\mathrm{relax}}$) and dephasing (described by $L_{\phi}$). The corresponding operators can be expressed in the following form:%
\begin{eqnarray} 
L_{\mathrm{relax}} = \sqrt{\mathrm{\Gamma}_1}\sigma^{+}, ~ ~
L_{\phi} = \sqrt{\frac{\mathrm{\Gamma}_{\phi}}{2}}\sigma_{z}
\end{eqnarray}
with $\sigma^{+} = \begin{pmatrix} 0 & 1 \\ 0 & 0 \end{pmatrix}$, $\sigma_{z} = \begin{pmatrix} 1 & 0 \\ 0 & -1 \end{pmatrix}$, $\mathrm{\Gamma}_1$ being the qubit relaxation, $\mathrm{\Gamma}_2 = \mathrm{\Gamma}_1/2 + \mathrm{\Gamma}_{\phi}$ is the  decoherence rate, $\mathrm{\Gamma}_{\phi}$ is the pure dephasing rate.
\begin{figure*}
	\includegraphics[width=1 \linewidth]{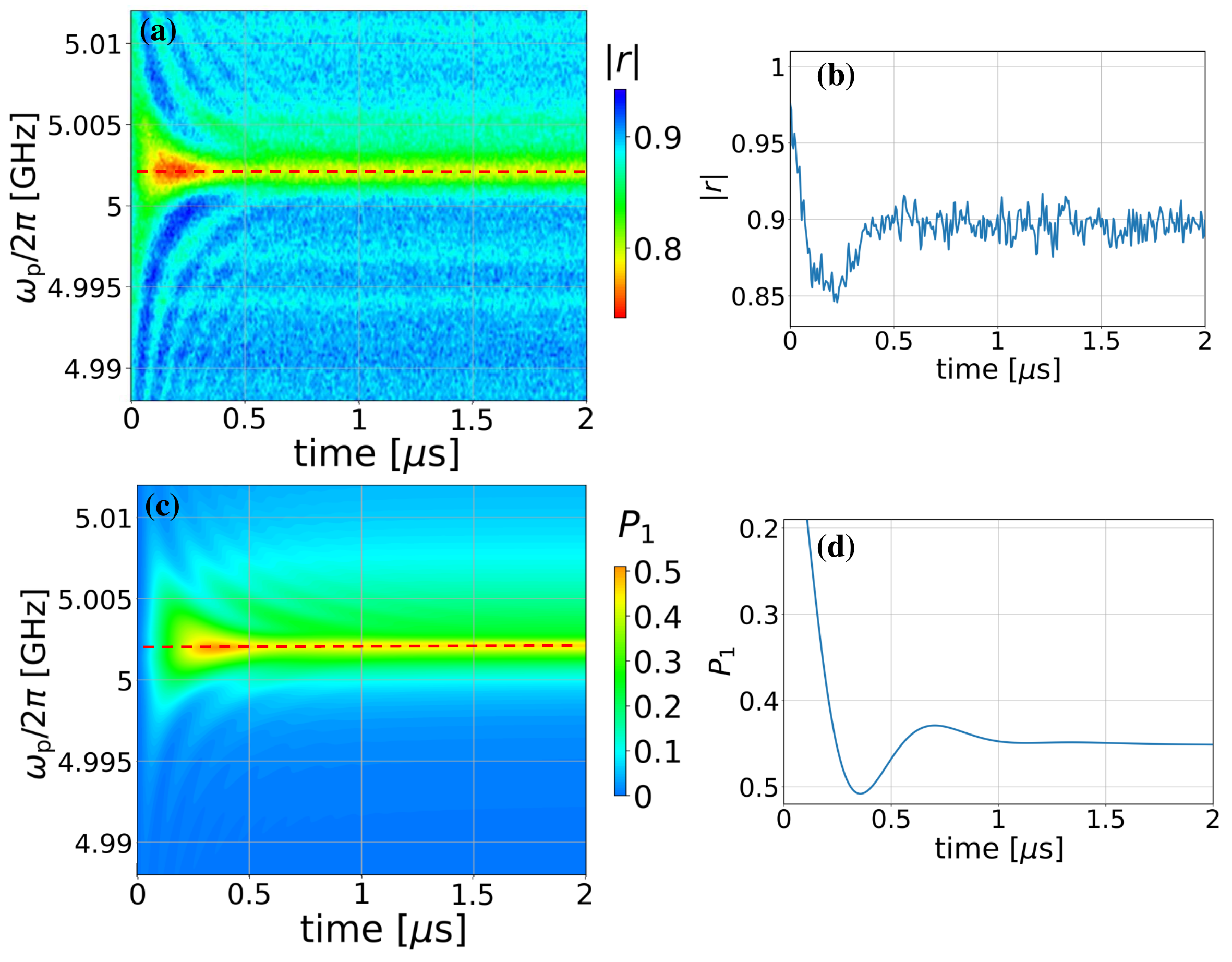}
	\caption{Coherent dynamics of the transmon qubit: dependence of the reflection coefficient $\abs{r}$ (the upper-level occupation probability $P_{1}$) with pump off on the probe frequency $\omega_{\rm p}$ and time $t$ for a weak probe $P_{\rm p}=\unit[-146]{dBm}$ [$G(\omega_{\rm p}/2\pi~=~5~\mathrm{GHz}) = 2\pi\times1.4~\mathrm{MHz}$] and $\omega_{10}/2\pi=\unit[5.002]{GHz}$. Plots (a, b) present experimental results; while (c, d) show plots computed theoretically. Panels (b) and (d) are line cuts of the experimental (a) and theoretical (c) plots, respectively, at $\omega_{\rm p}/2\pi=\unit[5.002]{GHz}$.
		\label{Fig_no_flux}}
\end{figure*}

\section{Interferometry and dynamics}
By solving Eq.~(\ref{Bloch_eq}) we obtain $P_{1}$ as a function of time $t$, pump frequency $\omega_{\rm pump}$, pump power $P_{\mathrm{pump}}$ (which corresponds to $\delta$ in theory), probe frequency $\omega_{\mathrm{p}}$, probe power $P_{\mathrm{p}}$ (which corresponds to $G$ in theory). The occupation probability is the function of all these parameters, $P_{1} = P_{1}(t,~ \omega_{\rm pump},~ \omega_{\rm p},~ \delta,~ G)$. The dependence obtained allows us to build, for instance, $P_{1} = P_{1}(\omega_{\rm p}, t)$. Also we can compute the dependencies for $P_{1}$ in a stationary regime by making the time averaging of the results. 

Figure~\ref{Fig_interferogram} shows a time-averaged interferogram, where $P_{1}$ is a function of $\delta$ and $\omega_{\rm p}$. We use the extracted parameters in Fig.~\ref{fig_calibration}, and select $G(\omega_{\rm p}/2\pi=5~\mathrm{GHz})=2\pi\times0.7~\mathrm{MHz}$ in Fig.~\ref{Fig_interferogram}; $G(\omega_{\rm p}/2\pi=5~\mathrm{GHz}) = 2\pi\times1.4~\mathrm{MHz}$ in Figs.~\ref{Fig_time_f_pump},~\ref{Fig_time_line},~\ref{Fig_time_delta},~\ref{Fig_no_flux} (the calibration between Rabi frequency and power, data is not shown) for the theory plots.  

For obtaining time-averaged values, we analyzed the curve $P_{1} = P_{1}(t)$ to extract the minimal time $t_{\mathrm{min}}$ after which the oscillation amplitude has no change. And then applied averaging for the interval $[t_{\mathrm{min}}, t_{\mathrm{final}}]$, where $t_{\mathrm{final}}$ corresponds to the time of the pulse turning off. We determined that for our case $t_{\mathrm{min}} = 1.5~\mu {\rm s}$ and $t_{\mathrm{final}} = 2.0~\mu {\rm s}$. 

Such interferograms are useful not only for obtaining the fitting parameters, but also they play a key role for characterizing the system. Particularly, such kind of figures:
\begin{itemize}
    \item allow to estimate the decoherence time of the system. Consider cases (a) and (b) of Fig.~\ref{Fig_interferogram}. We see that for the case (b) the peaks are separated, while for (a) they are not distinguishable. The maximal frequency $\omega_{\rm pump}$ for which we have a blurred picture (when individual resonances are not distinguishable) corresponds to the system decoherence time. So, one can conclude that $\Gamma_{2}/2\pi \simeq  1~\mathrm{MHz}$; 
    \item provide a tool for power calibration by interrelating the unknown distance between the zeros along the vertical axis in the experiment with the zeros of the Bessel function in theory; 
    \item provide novel opportunities for multi-photon spectroscopy. The resonances appear when $\omega_{\rm p} = \omega_{\mathrm{10}} \pm k\omega_{\rm pump}$, where $k$ is an integer number. Or, in other words, the system is resonantly excited when the dressed qubit energy gap is equal to the energy of $k$ photons, $k\hbar\omega_{\rm pump}$~[\onlinecite{Wen2020}]. 
\end{itemize}

In order to see the qubit dynamics we built the dependence $P_{1} = P_{1}(\omega_{\rm p}, t)$ for different pump frequencies $\omega_{\rm pump}/2\pi$~=~$5~\mathrm{MHz}$, $10~\mathrm{MHz}$, $15~\mathrm{MHz}$ in Fig.~\ref{Fig_time_f_pump}. As expected, for the stationary case, the resonances are observed at $\omega_{\rm p} = \omega_{\mathrm{10}} \pm k\omega_{\rm pump}$, and the value of the reflection coefficient $\abs{r}$ (occupation probability $P_{1}$) oscillates with period $T~=~2\pi/\omega_{\rm pump}$.

All theoretical plots were built by solving the Lindblad equation within the framework QUTIP (Quantum Toolbox in PYTHON) [\onlinecite{Johansson2012}, \onlinecite{Johansson2013}]. The function mesolve($H$, $\rho_{0}$, $c_{\mathrm{ops}}$, ...) from this library takes the Hamiltonian $H$ in a matrix form (in our case $H = H_{1}$), the initial state of the system $\rho_{0}$ (we assume that initially the system is in the ground state $\left|0 \right>$), the set of collapse operators $c_{\mathrm{ops}}$ which are related to the Lindblad superoperators (\ref{Lindblad_superoperator}) and some other parameters. The function mesolve($H$, $\rho_{0}$, $c_{\mathrm{ops}}$, ...) returns elements of the density matrix $\rho$ dependent on time.

\section{Conclusions}
 We considered the dynamics and stationary regime of a capacitively shunted  transmon-type qubit in front of a mirror, affected by two signals: probe and dressing (pump) signals. The multi-photon resonance dynamics, occurring at $\omega_{\rm p} = \omega_{\mathrm{10}} \pm k\omega_{\rm pump}$, consists of two temporal regimes: transient and stationary. In particular, we observed the dynamics of $k=0, \pm1, \pm2$ multi-photon resonances, because the node frequency is away from those resonances.  The occupation probability $P_{1}$ obtained with the Lindblad equation and the experimentally measured reflection coefficient $\abs{r}$ agree well with each other. Taking advantages of the strong coupling between the propagating field and qubit, and the ease of fabrication, superconducting qubits in front of a mirror provide a clear platform to study the dynamics of LZSM interference compared with other quantum two-level system [\onlinecite{Bogan2018}].   

\begin{acknowledgments}
I.-C.H.~acknowledges financial support from the Research Grants Council of Hong Kong (Grant No. 11312322). J.C.C. acknowledges financial support from the NSTC of Taiwan under project 110-2112-M-007 -022 -MY3 and 111-2119-M-007 -008. P.Y.W. acknowledges financial support from the NSTC of Taiwan under project 110-2112-M-194-006-MY3. M.P.L. and S.N.S. were supported by Army Research Office (ARO) (Grant No.~W911NF2010261). F.N. is supported in part by: Nippon Telegraph and Telephone Corporation (NTT) Research, the Japan Science and Technology Agency (JST) [via the Quantum Leap Flagship Program (Q-LEAP), and the Moonshot R\&D Grant Number JPMJMS2061], the Asian Office of Aerospace Research and Development (AOARD) (via Grant No. FA2386-20-1-4069), and the Foundational Questions Institute Fund (FQXi) via Grant No. FQXi-IAF19-06.
\end{acknowledgments}

\appendix
\section{Time dynamics of an atom-mirror system irradiated by probe and dressing (pump) signals}
Here we consider one more case of studying the time dynamics of the atom-mirror system irradiated by probe and dressing (pump) signals. 
One possible approach involves fixing the probe frequency $\omega_{\rm p}$ and analyzing the reflection coefficient $\abs{r}$ as function of time $t$, as shown in Fig.~\ref{Fig_time_line}. The measurements and calculations were done for various values of the pump frequency $\omega_{\rm pump}/2\pi~=~\unit[5]{MHz}$, $\unit[10]{MHz}$, $\unit[15]{MHz}$, with $\omega_{\rm p}/2\pi=\unit[5]{GHz}$. From the analysis of the plots one could conclude that: 
\begin{itemize}
    \item the probability and the reflection coefficient oscillate with period $T~=~2\pi/\omega_{\rm pump}$; 
    \item for the pumping frequency, $\omega_{\rm pump}/2\pi = \unit[5]{MHz}$, there are two kinds of peaks: high and low ones; 
    \item the system dynamics consists of two regimes: stationary and transient ones. The stationary regime is observed after $t$ = 1.5~$\mu {\rm s}$ for all the cases considered. 
\end{itemize}

Figure~\ref{Fig_time_delta} shows the dependence of the reflection coefficient $\abs{r}$ as a function of the probe frequency $\omega_{\rm p}$ and time $t$. The measurements and calculations were done for various values of the pump power $P_{\rm pump}$ ($\delta$ in the theory) and fixed pump frequency $\omega_{\rm pump}/2\pi=\unit[10]{MHz}$. From the plots we can deduce that increasing the pump power amplifies the resonances.

To understand better the influence of the pump signal on the system dynamics, we also show the plots with no flux pump. The corresponding results with $\delta = 0$ and $\omega_{10}/2\pi=\unit[5.002]{GHz}$ (in this case $\omega_{10}$ is slightly changed due to a slightly different flux bias) are shown in Fig.~\ref{Fig_no_flux}. Panel (a) shows the dependence of the reflection coefficient $\abs{r}$ on time $t$ and probe frequency $\omega_{\rm p}$, (c) is the corresponding theoretical result; panel (b) is the line cut of (a) at $\omega_{\rm p}/2\pi = \unit[5.002]{GHz}$, and (d) is the corresponding theoretical curve.

\nocite{apsrev41Control} 
\bibliographystyle{apsrev4-1}
\bibliography{references}

\end{document}